\documentclass[12pt]{iopart}
\usepackage{graphicx}
\begin{document}
\title{Prisoner's Dilemma with Semi-synchronous Updates: 
Evidence for a First Order Phase Transition}

\author{M Ali Saif} 
\address{Centre for Modeling and Simulation\\
University of Pune\\
Ganeshkhind, Pune, 411 007,\\
INDIA\\}
\ead{ali@cms.unipune.ernet.in}
\author{Prashant M Gade}
\address{Department of Applied Science,\\
 College of Engineering, Pune\\
Shivajinagar, Pune, 411 005\\
INDIA\\}
\ead{gade@unipune.ernet.in}
                                                
\begin{abstract}
Emergence of cooperation in self-centered individuals has
been a major puzzle in the study of evolutionary ethics.
Reciprocal altruism is one of
explanations put forward and
prisoner's dilemma has been a paradigm in this context.
Emergence of cooperation
was demonstrated for
prisoner's dilemma on a lattice with synchronous update [{\em{Nature, 
{\bf{359}}, 826 (1992)}}]. 
However, the cooperation disappeared for asynchronous
update and the general validity of the
conclusions was questioned
[{\em{PNAS, {\bf{90}}, 7716 (1993)}}].
Neither synchronous nor asynchronous updates
are realistic for natural systems. In this paper, we make a detailed study of 
more realistic system of 
semi-synchronous updates where $pN$ agents are updated at
every time instant.  We observe  a transition from all-defector
state to a mixed state as a function of $p$. Despite being transition
from absorbing state,  our studies indicate that it is  
a first order transition. Furthermore, we used damage spreading technique
to demonstrate that, the transition in this system  could be classified as
a frozen-chaotic transition.

\end{abstract}
\maketitle

\tableofcontents

\section{Introduction}

%Game theory has been successfully applied to several complex
%systems. Recently, physicists have shown growing
%interest in these applications as well \cite{szabo6}. 
%Tools of statistical physics have been used to analyze
%the behavior of such systems. The reasons for this kind of analysis
%are two-fold.
%One wants to gain insight in game-theoretical systems 
%from a variety of viewpoints. On the other hand, investigations in 
%a variety of systems and attempts to understand  
%their phenomenology, could enrich the overall theory of
%non-equilibrium statistical physics.
 
In ecological and social systems, game theoretical models
have been immensely successful and have emerged as a standard
models in certain cases. One of the interesting applications of game theory 
in these systems  has been explanation of emergence of co-operative
behavior. It was suggested that
cooperation could be sustained if there is repeated interaction and spatial
structure.
Prisoner's dilemma  on a lattice has emerged as a paradigm
in this  context \cite{axelrod}.
We will be analyzing a stochastic variant of the same
in this work.

"Prisoner's Dilemma" (PD) is an interesting problem in game theory where
cooperation between agents is strictly dominated by
defecting and the possible equilibrium is that everyone defects.
This is despite the fact that cooperation will lead to better 
average payoff for everyone. 
However, cooperation can arise in an iterated version of the 
game. This happens mainly due to threat of punishment for defection
in the next round.  If game theory
in traditional sense is applied to social system and 
if we assume common knowledge and perfect rationality for each agent,
cooperation is not a viable option.
Refining these assumptions makes cooperation possible. In iterated
games on a lattice, agents have knowledge only about their nearest neighbors
and the interaction is repeated. Cooperation is indeed observed
in these models. 
Thus it is claimed that the emergence of cooperation,  
which seems counterintuitive 
in the context of Darwinian evolution  but   observed
in several biological and social contexts, is  explained
by this model. 
In evolutionary games, the strategies are built in 
trial-and-error fashion and
more successful strategies are preferred in course of time.
This procedure leads to dynamics automatically. 
Obviously, societies comprise of several agents
who interact through a complex network of acquaintances. 
This fact has motivated the analysis of iterated version of 
PD on variety of networks. These studies have interesting consequences in
the social, biological and 
economic systems \cite{nowak1,turner1,turner2}. 
The choice of
underlying network depends on the problem being addressed.
In this work, we study the PD on a two dimensional lattice.
The reason is that evolutionary games on 2-D networks 
is a relevant and popular choice for several biological networks \cite{nowak2}.
(We must mention that recently 
other spatial 
structures have also received a fair share of attention \cite{szabo6,abramson,szabo1,vukov1,szabo2,
kim,vukov2}.) An extensive survey of modeling using game theoretical
models in ecology can be found in a recent book \cite{Novak}. In this paper, we will
focus on studying various phases in this system and nature of dynamic phase
transitions between these phases.

In this model,
the agents sitting on the nodes of two-dimensional lattice
update their strategies in a synchronous manner and payoffs
are computed using the payoff matrix of PD.
The major finding is that on introduction of spatial structure, 
there is emergence of  cooperative behavior 
among selfish individuals \cite{nowak2}.  However, this model was strongly
criticized by Huberman and Glance and also by Mukherji {\it {et al}}\cite{huberman, mukherji}.
They studied the robustness of these results with respect to
stochastic fluctuations and concluded that several
of the conclusions do not hold in presence of stochastic fluctuations.
One  stochastic variant studied in both these works has
been relaxation of requirement of synchronous update. 
In modeling physical phenomena, it is difficult to say whether
synchronous updating scheme is more natural than asynchronous scheme or
vice versa.  Models of traffic flow
have more realistic properties if updated 
in synchronous mode \cite{hinrichsen}.
However, if we are modeling physical phenomena in realm of 
equilibrium statistical physics, asynchronous updates offer
better results \cite{radicchi}. There could be phenomena where an intermediate 
mode could be more realistic.  In above model, certain degree of 
asynchronicity in updating strategies is relevant and possible and
this variant is studied in detail in this work.
We note that the 
differences between synchronous and asynchronous updates
have been a topic of recent interest in statistical physics community.
Apart from 
game theory \cite{blok}, it is studied in the contexts of 
boolean networks \cite{klemm,greil}, coupled map \cite{rolf}, 
neural Networks \cite{hopfield}, Monte Carlo processes \cite{choi} and biological 
networks \cite{klemm2,klemm3}. It has even been studied in the context of
equilibrium models \cite{radicchi} such as Ising Model.
In above work, nonequilibrium
phase transition, induced by introducing semi-synchronous
updating in Ising model, is studied. In the context of game theory,
there is a clear physical motivation for such study since it is
unlikely that even a strategy with slightly more payoff is
deterministically copied by everyone. 
Certain degree of stochasticity in updating of strategies is
indeed possible in these systems. Thus there is need for a detailed 
study of systems which do not evolve in a fully synchronous or
asynchronous manner.  
Hence, we make a detailed study of  the evolution
for semi-synchronous update where $pN$ members update
strategy every instant in PD on a lattice.

For completeness, we define the  2-person 
PD game in its classic form: this game describes
the confrontation between two players, each of whom may 
choose either to cooperate (strategy $C$), or defect (strategy $D$), at any
confrontation. If both players choose $C$, 
they get a pay-off of magnitude $R$ each; if one player 
chooses to $D$ while the other  chooses $C$, 
the defector player gets the biggest 
pay-off $T$, while the other gets $S$; if both players
defect, they get pay-off $P$. In this game, the pay-off values must
satisfy the inequalities $T>R>P>S$ and $2R>S+T$. For such choice
of parameters, the paradox is evident. Each player is tempted to 
defect, but they would be worse off if both 
defected and total payoff for both together would be 
higher if they co-operated in stead. However, best
payoff for an individual player is obtained when he
defects while the other player cooperated.

In ecological context,
Nowak and May simulated this system with choice of 
parameters $R=1$, $T=b$ $(1.0<b<2.0)$ and $S=P=0$ \cite{nowak2,nowak22}. 
They believe that most of the interesting behavior is reproduced
for this choice of parameters. They studied 
the PD on a two dimensional array 
with synchronous updating. They explored 
the asymptotic behavior  of the game for various values 
of the $b$. Here  players interact with their 
local neighbors through simple deterministic rules and 
have no memory of past. They found that, the dynamical
behavior of the system depends on the parameter b.  
For a range of values of $b$ $(1.0<b<2.0)$, system reaches 
a steady state with a non-zero fraction of cooperators.
They concluded that spatial structure and repeated interactions
promote cooperation in the PD.

This synchronous updating came in for heavy criticism. 
It was argued that, the global clock in the social
systems are very rare and the probability of two events
evolve exactly at same time has measure zero. It  
was also argued that more realistic modeling would involve updating 
the system by individual. Thus only one player should be updated at 
each time step. This type of updating is
called asynchronous update. As we mention previously, 
Huberman and Glance studied
PD game with same parameters of Nowak 
and May under the asynchronous updating rule \cite{huberman}.
A similar argument was made by Mukherji {\it et al.} \cite{mukherji} 
and Masaki and Mitsuo \cite{tomochi}.
In this case, the system rapidly converges 
to steady state where all the players become defectors. 
They argued that the previous results about emergence of 
cooperation  are not generic. 
Nowak et al \cite{nowak7,nowak4,nowak3}, replied that if they study the
behavior in entire parameter space, cooperation is
observed even for asynchronous update for some choice
of parameters.
They found that, the two updating rules are similar for a 
some  values of $b$, but for $(1.8<b<2.0)$ the two
updating rules lead to different steady states.
They also argued that 
discrete time is appropriate for many 
biological situations where interaction phase is followed 
by reproduction phase. Thus synchronous update is more
relevant and realistic biologically. However, neither completely
synchronous nor completely 
asynchronous updating is realistic in natural processes and there
are bound to be stochastic fluctuations in updating.  There 
have been attempts to interpolate 
between these two cases \cite{radicchi,blok,klemm,greil,rolf,hopfield,choi,klemm2,klemm3}. 
In this work, we make a detailed study of PD on a 
2-D lattice with semi-synchronous updates from the 
viewpoint of statistical physics and investigate how generic
the results are. 

We will make a detailed study of the observed
phases for semi-synchronous update and present a
phase diagram. The phases of interest in this system are 
an all-defector state and a mixed phase with cooperators and defectors.
In particular, we will study the transition between these two
phases as a dynamic phase transition.

In statistical physics, a lot of effort is devoted
for finding the order of transitions and critical exponents in
case of continuous phase transitions.  The reason is that the critical 
behavior lets us distinguish between essential and not
so essential details of the system. Idea of universality 
in the theory of phase transitions has allowed us
to see how seemingly disparate models have common
underlying features. Thus it is important to study the transitions in detail.  
All agents becoming $C$ or $D$, is an absorbing state while
coexistence of $D$ and $C$ can be considered as an active phase. 
Hence, this system characterized by two absorbing state, all defectors state and all cooperators state.
It has been long argued for that all one component systems 
undergoing from active phase transition to a unique absorbing state have a phase
transition in the class of directed percolation (DP) if the order parameter
is a scalar and there are no extra symmetries or 
conservation laws and the interaction through short-range \cite{hinrichsen}.  Also, Most system with multiple
absorbing state found to fall in the DP class \cite{marques}. 
We will study the veracity of this conjecture.
Firstly we need to establish the order of the transition.
The first (second)  order phase transitions  in equilibrium systems 
are characterized by discontinuities in the first 
(second) derivatives of free 
energy e.g., the internal energy and order parameter. This singularities 
at a first order phase transition are due to phase coexistence and there 
are no critical exponents.

For PD game system, it was found that, for different choice of updating rule
transition to all-defector state on variation of parameter $b$ is in the 
class of directed percolation \cite{hauert,guan}. We will study the Nowak
and May's 
system under the variation of probability $p$ of update. 
{\em{We observe that the transition is not continuous.}}
We must also mention that there are several known exceptions 
in nonequilibrium systems
which exhibit a discontinuous
transition to absorbing state \cite{dickman1,bagnoli,adam,evans,monetti3,cardozo,oerding}.

Our updating strategy is as follows. We allow  every 
player to update his strategy
 with probability $p$ in each Monte Carlo step. Under variation of 
this probability from $p\rightarrow 0$
(asynchronous) to $p=1$ (synchronous) in thermodynamic
limit, we study the effect of the updating 
scheme on the behavior of the system. 
In second section of the paper, we establish that the
transition is indeed a first order transition.
In the third section, we  carry out a  
damage spreading analysis to study the dynamical phase digram and critical 
behavior. 

\section{The Model and Simulation}
We investigate  PD on 
the two dimensional lattice of size $L$ with evolutionary dynamics. 
The agents on each site of lattice 
can choose  only two strategies ${D=0,C=1}$. $D$ corresponds to  
defector while $C$ corresponds to  cooperator. 
(The defectors and cooperators have 
also been viewed  as dead and living sites in some applications.) 
We employ fixed boundary conditions and assume that the agents 
have no memory of the past. Initial configuration consists of 
$30\%$ defectors and $70\%$ cooperators distributed randomly  
on the lattice. (We checked other 
initial conditions as well. We varied the density of defectors 
between  $10\%$ to $50\%$ and found that the asymptotic stationary state
did not change.) Every agent interacts with  eight nearest neighbors  
and self.  We set the parameters $ T=b (b=1.83)$, $R=1$ and $S=P=0$.
 The pay-off matrix is\\
%\begin{table}[htbp]
%	\centering
\begin{center}
		\begin{tabular}{c|c|c}
%			\begin{align}
	&C&D\\\hline
	C&1&0\\\hline
	D&b&0\\
%\end{align}
		\end{tabular}\\
		\end{center}
%\end{table}

The players interact simultaneously and independently 
of each other. Their pay-off is the sum of the 
pay-offs from all nine interactions (with neighbors and self). 
Generally, each player updates its strategy by imitating the strategy of 
most successful agent in the neighborhood.
The main variation in this work is as follows: At every time-step,
each player updates his strategy by adopting the strategy of the 
most successful neighbor {\em{with probability $p$}}. 
For asynchronous update, only one agent
updates strategy at every instant. This could be compared with
evolution with $p=1/N$ where $N$ is total number of agents.
On the other hand, all agents update their strategy for 
synchronous update and the above rule for $p=1$ is same
as synchronous update which is widely studied.
We vary the value of $p$ from $p\rightarrow 0$ 
to $p=1$ interpolating between asynchronous and synchronous update.
The phases of interest are all-defector and mixed phases. 
We study the domains in $p-b$ plane, where almost all initial 
conditions lead to one of these phases.

We carried out a detailed investigation of this system for
parameter values in the range $0.0<p<1.0$ and 
$1.0<b<2.0$. The corresponding phase diagram is shown in Fig. 1.
For values of $b$ in the range $1.8<b<2.0$, the final phase 
depends on the value of $p$.  In this work, we study the nature of the phase 
transition observed  on varying $p$.
The density of cooperators $\rho_c$ is an obvious order parameter for
describing the transition since it is zero in an all-defector
state and positive for mixed state.
\begin{figure}
 \includegraphics[width=70mm,height=60mm]{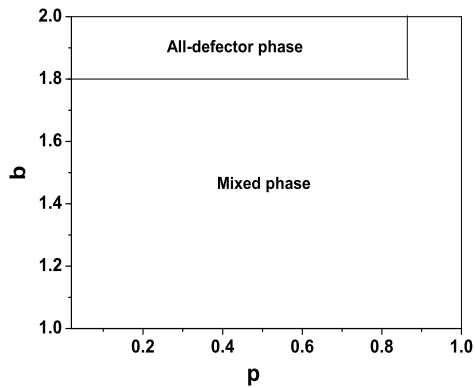}
 \caption{\label{fig1:}Schematic phase diagram of the semi-synchronous PD game. 
We plot the phases observes as a function of $p$ (probability of update for 
any given site) and  $b$ (temptation to defect).
For the values of $1.8<b<2.0$ system switches from absorbing state (all defectors) 
to an active state where cooperators and defectors coexist.
Simulations are carried out for $L=60$, 
and initial 1000 time-steps are discarded. We average over
100 different initial conditions.}
\end{figure}

We compute the value of $\rho_c$ in steady state
for different values of updating parameter $p$. We also investigate 
the same for  different lattice sizes $L$. We observe that,  when
we change  $p$ from $p\rightarrow 0$ to $p=1$, the 
system switches from an all-defector  absorbing state with $\rho_c=0$ 
to active phase with $\rho_c >0$. We find a clear evidence
of  the metastable state near the transition region.  In Fig. 2, we plot 
the lifetime (average time taken by the system to 
reach an all-defector absorbing state) as a function of
updating probability $p$ for 
several values of  $L$. For smaller values of lifetime
($<10^6$), we average over $100$ 
configurations and for larger lifetimes, we average over 
$10$ configuration. Depending on the 
value of the parameter $p$ we can distinguish between three regions:
 
\begin{itemize}
\item For a long range of the values of the parameter $p<p^*$,  
the system rapidly converge to absorbing all-defector state.
For $p> p^*$ the time required to reach this absorbing state grows
abruptly compared to $p$ values smaller than $p^*$ Fig. 2.
In this region, the lifetime
approximately equally for all values of parameter $p$ and equal to 
the lifetime in asynchronous update. Thus one could say
that the behavior of system is analogous to one obtained in
purely asynchronous updates.  As shown in Fig. 2, the value $p*$ 
converge to finite value at thermodynamic limit.  
\item There is very sharp range of the value of parameter $p^*<p<p^{**}$
where 
the system falls into metastable state. The mixed state in which both
cooperators and defectors are present is extremely long-lived and
system falls in an all-defector state after a very long time.
The average time taken by the system to reach this absorbing
state increases with updating probability $p$ till it reaches a very large 
value as
$p \rightarrow p^{**}$. For any value $p>p^{**}$, the system remains in the active phase. 
As $p \rightarrow p^{**}$ the system spend longer time in 
the active phase before it collapse eventually to its absorbing state. We 
found that, for the best fit the lifetime as a function of
$(p^{**}-p)$ is an exponential decreasing fit.   (See Fig. 3)
\item For values of $p>p^{**}$ , we indeed found that the system  saturates 
with finite number of cooperators. We would like to 
assert that, as the system crossover the metastable state region
toward this region the order parameter $\rho_c$ exhibits a certain
jump in its value (See Fig. 4). 
\end{itemize}

%figure1
 \begin{figure}
 \includegraphics[width=70mm,height=60mm]{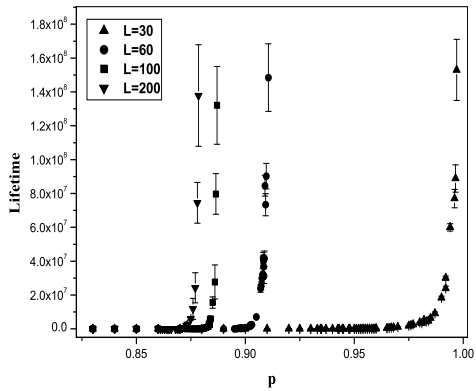}
 \caption{\label{fig2:}The average time required for the
system  to reach its absorbing state (lifetime of coexistence 
state) is plotted as function of updating probability $p$ for different 
lattice sizes. The lifetime changes 
abruptly and diverges near the critical point.  }
 \end{figure}
 
 %figure2
 \begin{figure}
 \includegraphics[width=70mm,height=60mm]{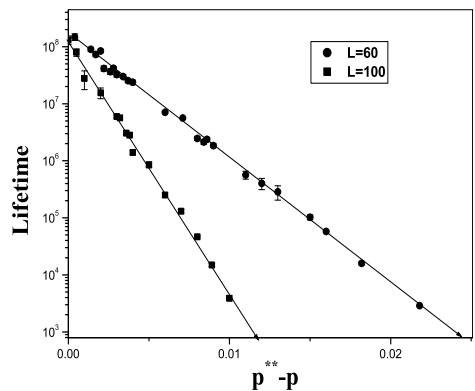}
 \caption{\label{fig3:}The semi-log plot for the lifetime
of coexistence stat as a function of  $(p^{**}-p)$ is plotted 
 for $L=60$ and $100$ in the metastable state. ( We note that
$p^{**}$ is the value of updating probability $p$ above which system reaches
a saturated state with coexistence of cooperators and
defectors.) The solid line shown the exponential fit for the data.}
 \end{figure}
 %end2 

Appearance of metastable state in this system shown us
some similarity to equilibrium discontinuous transition. 
It is known that, the discontinuous transition accompanied usually by 
metastability. 

To confirm that for any value of $p<p^{**}$ the 
system collapses to an absorbing state, we plot
the time evolution of density of 
cooperators $\rho_c(t)$ as a function of time $t$ in
Fig. 4. We show the behavior of $\rho_c(t)$ as a function of 
$t$ for various values of $p$ for a system of  
size $L=60$. We average over $10^3$ different initial conditions. The
finite number of cooperators are observed for some time
(depending on the value of $p$. The time taken to reach absorbing state 
increases exponentially as $p \rightarrow p^{**}$ as  mentioned 
previously) followed by a collapse to an  absorbing state. 
As $p \rightarrow p^{**}$, the curves in Fig. 4 
becomes flatter (the system need longer time to approach absorbing state) 
till the system reaches its steady state at
$p>p^{**}$. Fig. 4 shows  that, when the steady state of the
system changes from an all-defector state to coexistence state,
there is certainly a jump in the value 
of the order parameter $\rho_c$. Metastability and long time
required by the system to reach its steady state make it very difficult
to locate the critical value of $p$ above which the system reaches
its stable active phase. We run the programs for very long times
so that correct estimates can be made.
In the Fig. 5, we present the order parameter $\rho_c$ 
as function of the parameter $p$. We used lattice of size $L=100$, 
we averaged over $10^2$ samples after discarding $10^6$ transient time. 
The figure shows the clear jump in the value of order parameter $\rho_c$.     
 
This result of a clear jump in the order parameter value at the
transition point coupled with a presence of 
metastable state indicate  that the model undergoes a first order phase
transition between the active and absorbing phase.

%figure3
\begin{figure}
\includegraphics[width=70mm,height=60mm]{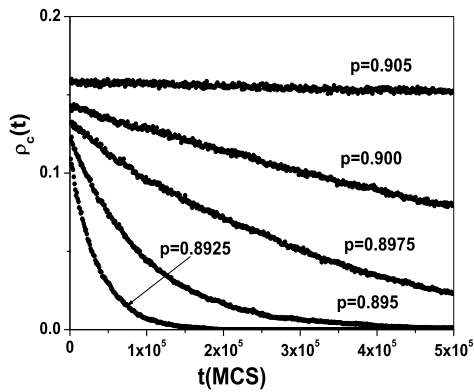}
\caption{\label{fig:4} The time evolution of the density
of cooperators $\rho_c(t)$ for $L=60$ for different values of 
the parameter $p$ near the steady state region. 
On increasing value of $p$  from $p=0.8925$ (bottom curve) 
to $p=0.905$ (top curve) system needs more time to reach 
absorbing state. It's clear that at $p=0.905$,
 the curve becomes flat, and system reaches the active coexistence
phase asymptotically.  We start simulation with $70\%$ 
cooperators in all cases. There is a clear jump in the asymptotic 
value of $\rho_c(t)$.}
\end{figure}
%end3 

\begin{figure}
\includegraphics[width=70mm,height=60mm]{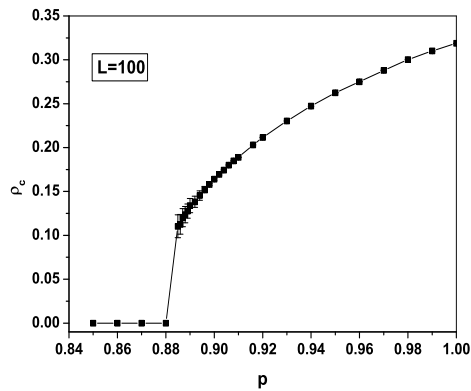}
\caption{\label{fig:5} The asymptotic value of order parameter 
$\rho_c$ as function of $p$ for lattice size $L=100$. 
A clear jump in the value of order parameter at the 
critical point indicates a first order phase transition.}
\end{figure}

\subsection{Analysis using Binder cumulant}
As mentioned above, presence of long-lived
metastable states makes it very difficult to locate the critical point
and the kind of phase transition since one could always doubt if
the jump in order parameter is true.
Fortunately, fourth-order reduced Binder cumulant 
offers a  precise tool 
which is very sensitive to the nature of the 
phase transitions. The fourth-order reduced  Binder cumulant of the 
order parameter $\rho_c$ is defined \cite{binder1,binder2} as:\\
\begin{equation}
U_L=1-\left\langle \rho_c^4\right\rangle/3\left\langle \rho_c^2\right\rangle^2,
\end{equation}
Systematic analysis of Binder cumulant $U_L$  
has been successfully used to determine the 
order of the phase transition in several equilibrium phase transition 
\cite{bahiana2,koiller,ferrenberg,mccarthy,yin,acharyya,
szolnoki,hasmy,challa,lee,zhang}. In the second order phase transition,
$U_\infty=2/3$, at the transition point in 
the thermodynamic limit. On the other hand,  for first order transitions,
it has a minimum at transition point. The reason is as follows.
For a continuous phase transition, the distribution of values
of order parameter is always a Gaussian, the position of which keeps
changing. For a first order transitions, distribution is
different. Here, we have phase coexistence and distribution
is superposition of two Gaussian centered at values corresponding to
each phase \cite{binder1,binder2,bahiana2}.
The quantity $U_L$ is not well defined numerically when
$\rho_c \rightarrow 0$ which is the case here. 
To overcome this difficulty we follow \cite{bahiana2,koiller} and 
add an arbitrary fixed constant to all values of the order parameter $\rho_c$, 
(we fixed it to be equal $0.001$ in this work), that is rigidly
shift the probability distribution of the order parameter away from zero.
 
We have plotted Binder cumulant $U_L$ as function of probability $p$ 
for various $L's$ in Fig. 6. We average over $150$ configurations
and $5\times10^6$ iterations on discarding $10^4$ transients. A 
clear minimum in the value $U_L$ shows that the transition is of first order. 
The value of probability $p$ that corresponds to a minimum of the Binder 
cumulant $U_L$ in the thermodynamic limit is the critical point.\\
%figure5
\begin{figure}
\includegraphics[width=70mm,height=60mm]{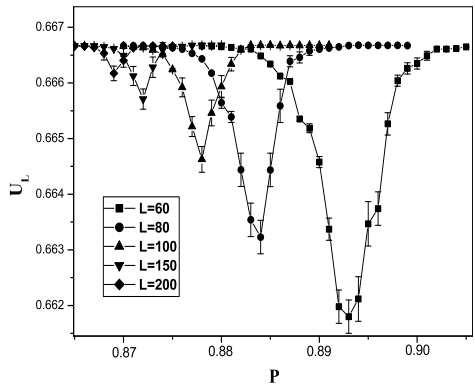}
\caption{\label{fig:6} Binder cumulant $U_L$ is plotted as function
of the probability $p$ for different lattice sizes. The 
presence of a minimum suggests that the transition is 
of first order.}
\end{figure}
%end5
It is well known that for  the first order phase transition, finite-size 
scaling theory predicts rounding and shifts of the critical point to be
inversely proportional to the volume, $L^{d}$ in $d$ 
dimensions \cite{binder1,binder2,peczak}. We have plotted 
$p_c(L)$ as function of $L^{-2}$ in Fig. 7. We found an 
excellent linear behavior which matches with finite- size scaling prediction. 
In the thermodynamic limit the value of $p_c=0.8678(6)$ for the best fit.

%figure6
\begin{figure}
\includegraphics[width=70mm,height=60mm]{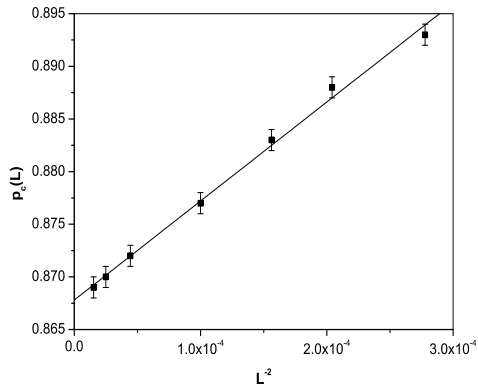}
\caption{\label{fig:7} The critical point $p_c(L)$ is plotted
as a function of 
$L^{-2}$. The data came
from the same simulations results as in Fig. 6. An excellent linear
behavior matching with finite-size scaling predictions
for first order transition is obtained.}  
\end{figure}
%end6

\subsection{Hysteresis}
Hysteresis effect is a characteristic feature of first order phase
transitions. The magnetic systems show a first order transition on
varying magnetic field but a second order transition
on varying temperature. Thus there is a hysteresis on variation of 
magnetic field, but not on varying temperature. Thus
is an useful tool to distinguishing first order phase 
transition  from continuous  phase transition.
The reason for hysteresis in first order phase transition is
due to the coexistence
of two phases. However, there is one difficulty in 
studying hysteresis in transitions leading to
an absorbing state. If the system falls in an absorbing state, it can not
come out. Thus we need to suitably modify the model.
Hence, we use the spontaneous creation 
method (SCM) \cite{bagnoli,bidaux,monetti1,dickman}. 
The SCM overcomes this difficulty by allowing for small nonzero $\epsilon$ 
concentration 
of active site to survive. For the second order phase transition, this 
spontaneous creation rate of active sites will destroy the phase transition. 
However, for first order phase transition a small spontaneous creation 
does not change the nature of the transition. It only turns the absorbing 
state into a fluctuating state of average density $\epsilon$.

We carry out simulations for 
$L=200$ and allow $\epsilon=0.0005$ fraction of active sites to
survive. We  vary $p$ stepwise in steps of $\Delta p=0.005$.
We record the value of the density of cooperator $\rho_c$ after $t_r$ update. 
We average over $100$ loops and plot the density of cooperators 
$\rho_c$ as a function of $p$ in Fig. 8, for different relaxation time $t_r$. 
The evidence of 
hysteresis effect confirms our conclusion that the transition in the
above system is a first order transition.
\begin{figure}
\includegraphics[width=70mm,height=60mm]{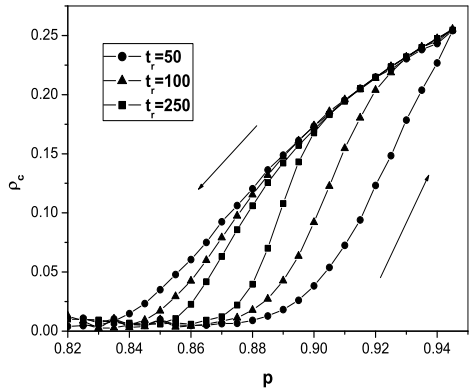}
\caption{\label{fig:8} Hysteresis loops of the density of cooperators 
are obtained 
by SCM method for lattice size $L=200$. They are obtained  for different relaxation 
times $t_r$ and it is clear that the loop area decreases for higher $t_r$. 
The loops are generated counterclockwise.}
\end{figure}

\section{Damage Spreading Analysis}

The damage spreading technique was suggested for 
first time by Kauffman in the context of 
biological systems \cite{Kauffman}. Recently, This concept attracted much 
attention and applied to analyze several dynamical systems such as  
cellular automata \cite{atman}, kinetic Ising models \cite{hinrichsen1} and 
surface growth model \cite{kim2}. 
%to name a few.

In the damage spreading technique, we 
follow the time evolution of two almost identical 
configurations. The initial condition for the second configuration
is same as the first configuration except 
for small perturbation. Now one studies the evolution of
 these configurations in time under the same dynamics  and see
how the initial perturbation (damage) propagates.
In our case, we start the simulation  system from  a
random initial configuration. We allow the configuration to evolve 
until reaches its steady-state. Let us label this 
configuration as $\sigma^A$ (first copy). The 
$\sigma^B$ (second copy) is created 
from saturated state of the 
first copy $\sigma_A$ by carrying out a small perturbation or damage 
to the this copy. Now both the copies are evolved under same dynamics
in the following sense. We update the strategy of any agent in system A
with probability $p$. Whenever we choose to update (not to update)
the strategy of
agent in the first copy, we also update  (do not update)
the strategy of the agent in second copy.  This synchronized updating
ensures that same set of random numbers is used during updating.
We study the time evolution of both configurations under this dynamics. 
Evolution of the  configuration can be described by
trajectories in the phase space.  The question is how the difference between 
these initial conditions grows or decays as a function of time. In other
words, the question is if the damage heals or spreads.
If the 
two initially close trajectories quickly become different, it is generically 
called chaotic. 
In order to measure the difference between that two 
systems, a useful metric is given by the Hamming 
distance or damage define by \cite{herrmann}
\begin{equation}
D(t)=1/N \sum\limits_{i=1}^N \left| \sigma^{A}_i(t) - \sigma^{B}_i(t)\right|
\end{equation}
where $N$ is the number of sites of the system. The quantity $D(t)$  measures 
the fraction of sites of configuration $\sigma^A$ which have 
different strategy from those from configuration
$\sigma^B$. In the thermodynamic limit, $D(t)$ may go to zero
if damage heals completely while it may tend to a positive value
if it does not heal. For chaotic system, we expect the damage
to grow in time and reach certain asymptotic value.
However, in the 
so-called frozen phase $D(t)$ will go to zero \cite{albano1,albano2,rieger}.

Nowak and May, in their original paper, do not 
exactly make a systematic study of damage spreading as done in present
work. However, they empirically simulate different configurations and
call a phase chaotic if a slightly perturbed initial condition
leads to a very different state. They state that
the steady state of PD with
synchronous update is chaotic for the parameter range $1.8<b<2.0$ \cite{nowak2}. 
This is very similar to the definition of chaotic phase in damage
spreading studies.
In this work, we will make a systematic study of damage spreading in
PD as a function of updating probability $p$.
We would like to mention that similar studies were carried out 
for the Stochastic Game of Life (SGL) by Monetti and Albano. They found
a first order transition as a function of stochasticity and a
rich dynamic critical behavior in the system\cite{monetti2}. 
Since our system is similar, we expect an
analogous dynamic behavior in our system.

\subsection{Analysis of Damage Spreading}
We followed the same updating role and initial values as 
described in the previous section. We start our 
simulation with 
$70\%$ cooperators and $30\%$ defectors distributed randomly on the 
sites of square lattice. As mentioned before, we update the strategies
with probability $p$ and wait till
the system reaches its steady state and we label this 
configuration as $\sigma^A$. The second configuration $\sigma^B$ 
created with small damage in the central sites of configuration $\sigma^A$. 
The evolution of damage spreading is computed as a function
of time for different update probabilities. 
We have plotted $D(t)$ as a function of $t$
in the active phase for various values of $p$ in Fig. 9(a) In the log-log scale 
the damage $D(t)$ evolves linearly as function of $t$ until it 
saturates to a  finite value of damage after some time. 
In the Fig 9(b), we have plotted the average asymptotic value  of 
damage $\left\langle D(\infty) \right\rangle$ for different values of the 
parameter $p$ for lattice of size $L=200$. It is clear that
below a certain threshold probability, 
the  damage reaches zero asymptotically. This threshold probability 
is very close to $p_c$ of the transition mentioned above.
So, we conclude that the dynamical phase transition behavior in this system 
between the absorbing all-defector state and 
mixed state is intimately connected to  frozen-chaotic 
transition in damage spreading.

\begin{figure}
\includegraphics[width=70mm,height=60mm]{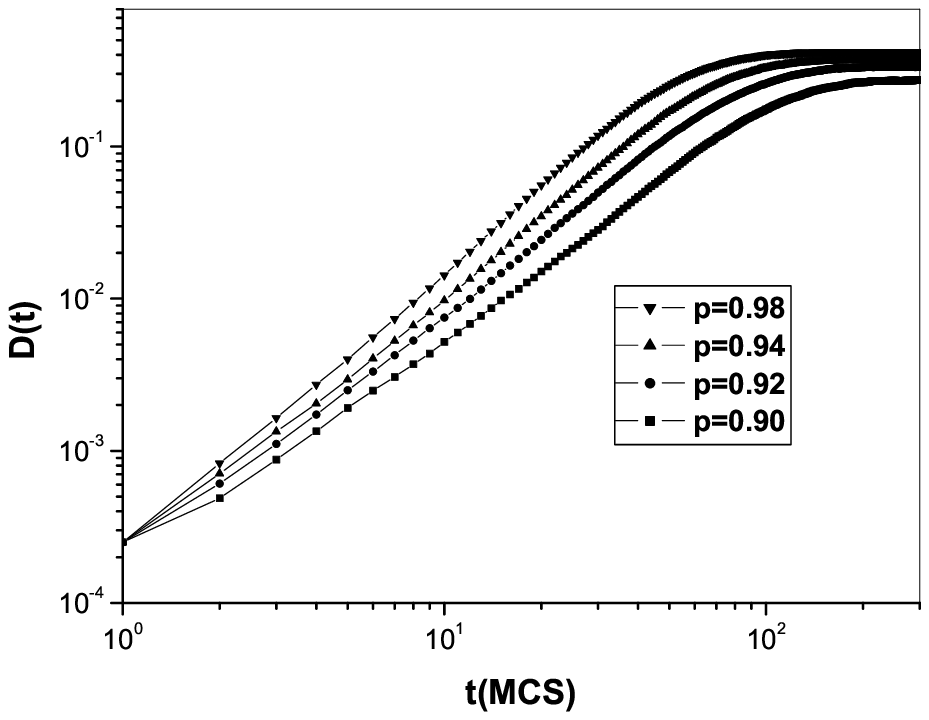}
\includegraphics[width=70mm,height=60mm]{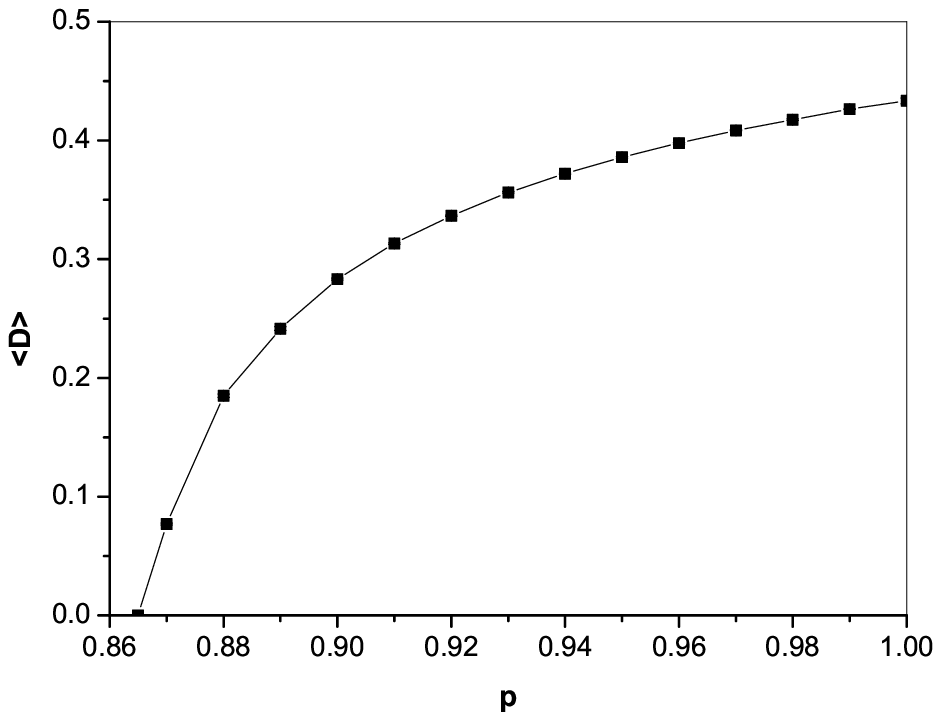}
\caption{\label{fig:9}(a) Damage spreading $D(t)$ is plotted as a function of
time for lattice 
size $L=200$ for different values of parameter $p$. We average over $100$ 
configurations. (b) The average asymptotic value of damage $ \left\langle D \right\rangle$ 
for different values of the parameter $p$ for lattice size $L=200$}
\end{figure}

We also study the effect of finite lattice size on
damage spreading.
In Fig. 10 (a), we plot the damage $D(t)$  as function of $t$ 
for various value of lattice size $L$ for $p=1$. 
The behavior obtained is  very similar to one observed 
in SGL \cite{monetti2}. It is the evident that, the slopes and saturation 
values of all curves is independent of lattice size $L$.
However, the time required for the system to reach the plateau $\tau(L)$ 
increases with $L$.  The behavior can be summed by the
scaling form \cite{monetti2}

\begin{equation}
D(x)\propto \left\{ \begin{array}{ll}
x^\alpha &\mbox{$x\leq 1$};\\
const& \mbox{$x>1$}.\end{array} \right.
\end{equation}

where $x=t/\tau(L)$ and $\alpha$ is an exponent. In Fig. 10(b), we
show the scaled data.  It is clear that the data for 
four different lattice sizes collapses to a 
single curve using above scaling form.
We calculate the value of $\alpha$ for the different value of the 
parameter $p$ in the Table 1 .
\begin{figure}
\includegraphics[width=70mm,height=60mm]{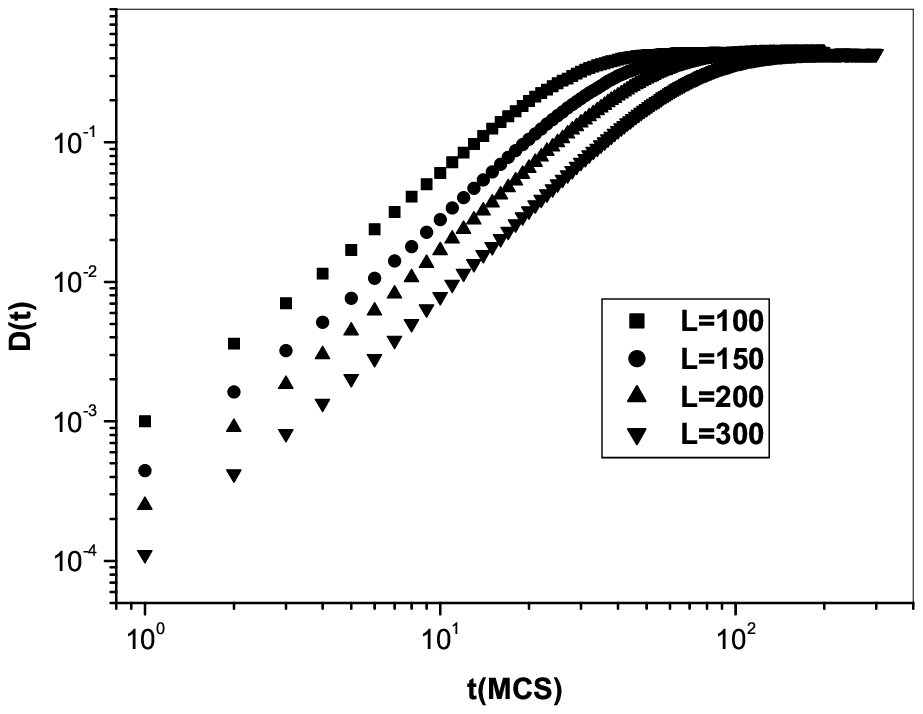}
\includegraphics[width=70mm,height=60mm]{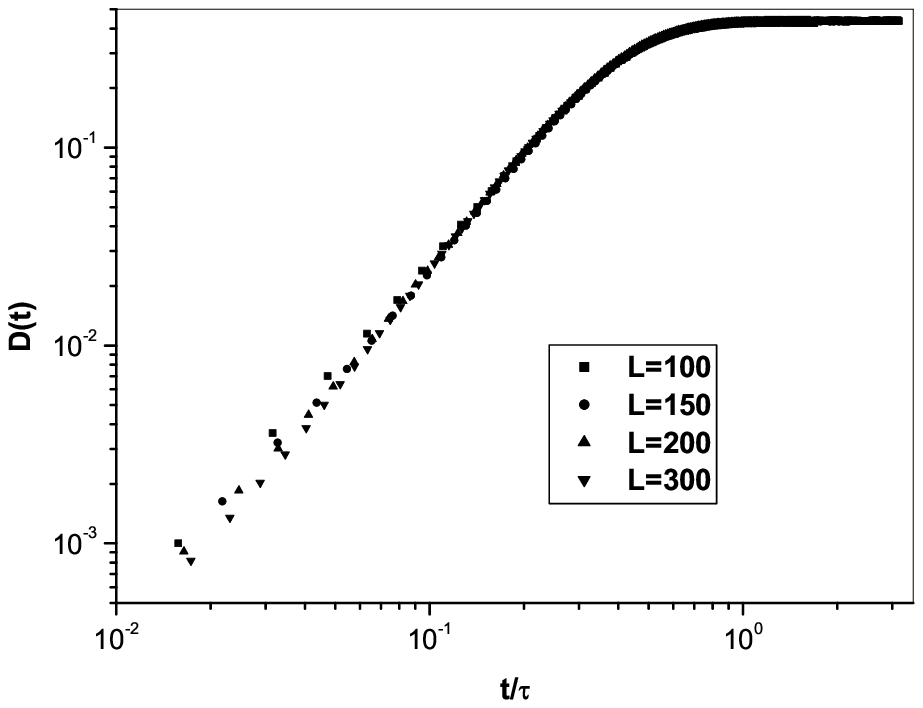}
\caption{\label{fig:10}(a) Damage spreading $D(t)$ as function of $t$ 
obtained for different lattice size $L$ at fixed value of the 
parameter $p=1$. (b) The damage $D(t)$ versus $t/\tau$ for the data 
shown in (a).  The data for four different lattice size collapse 
onto a single curve.}
\end{figure}

The average mean-square distance $R^2(t)$ over which the initial damage 
spreads from the center of the lattice toward the boundary is 
also calculated in following manner.  We start the 
simulation at $t=0$ with initial damage of one site at
the center of the lattice. 
Lattice size is fixed at $L=199$ and  
mean-square distance $R^2(t)$ is computed as a function of time $t$.
We carry out the simulations  
for various values of parameter $p$. We average over 100 configurations and
the results are plotted in Fig. 11.
It is found that $R^2(t)$ and $D(t)$ exhibit a similar behavior. 
Thus the following scaling ansatz should  hold \cite{monetti2}

\begin{equation}
R^2(x)\propto
\left\{ \begin{array}{ll}
x^\beta &\mbox{$x\leq 1$};\\
const& \mbox{$x>1$}.\end{array} \right.
\end{equation}

where $\beta$ is an exponent.
The value of exponent $\beta$ is tabulated as a function of $p$ on
Table 1.

\begin{figure}
\includegraphics[width=70mm,height=60mm]{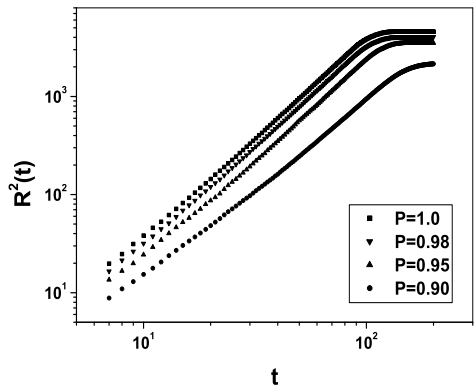}
\caption{\label{fig:11}The average mean-square distance for epidemic
spreading $R^2(t)$ is plotted as a function of $t$ for lattice size $L=199$ 
for different values of  $p$.}
\end{figure}

The number of damaged sites is related to the spatial extent of the 
damage through the fractal dimension $d_f$ of the damaged cloud, so
\begin{equation}
D(t)\propto R^{d_f}(t)
\end{equation}

Then from the Eqs. (3) and (4) we get that
\begin{equation}
2\alpha=\beta d_f
\end{equation} 
   
\begin{table}
\begin{center}
\caption{\label{Table1}Values of the exponents $\alpha$,$\beta$ 
and $d_f$ at the different value of the parameter $p$.}
\begin{tabular}{cccc}
\hline
\hline 
$p$ & $\alpha$& $\beta$& $d_f$\\
\hline
1.0&1.96(9)&2.03(7)&1.93(1)\\
0.98&1.88(7)&2.01(5)&1.87(1)\\
0.96&1.85(8)&1.98(9)&1.86(8)\\
0.92&1.64(2)&1.90(1)&1.72(6)\\
0.90&1.52(8)&1.82(2)&1.67(1)\\
0.88&0.97(8)&1.59(4)&1.22(1)\\
\hline
\hline
\end{tabular}\\
\end{center}
\end{table}

We tabulate values of $d_f$   in 
Table \ref{Table1} using Eq. 6. 
While the fractal dimension $d_f$ is found very close to 
$2$ for synchronous update($p=1$), it deviates appreciably for smaller
values of $p$.    
For the value of  $p$ near the critical point $p_c$,
the fractal dimension $d_f$ is smaller. We find that
the damage spreading 
near the critical point is certainly a fractal object. 
On the other hand, the snapshot in Fig. 12 shows 
that for synchronous update ($p=1$),
the damage spreading is approximately compact. Thus
the damage spreading is fractal near the critical point
and compact for synchronous update. 

\begin{figure}
\includegraphics[width=70mm,height=60mm]{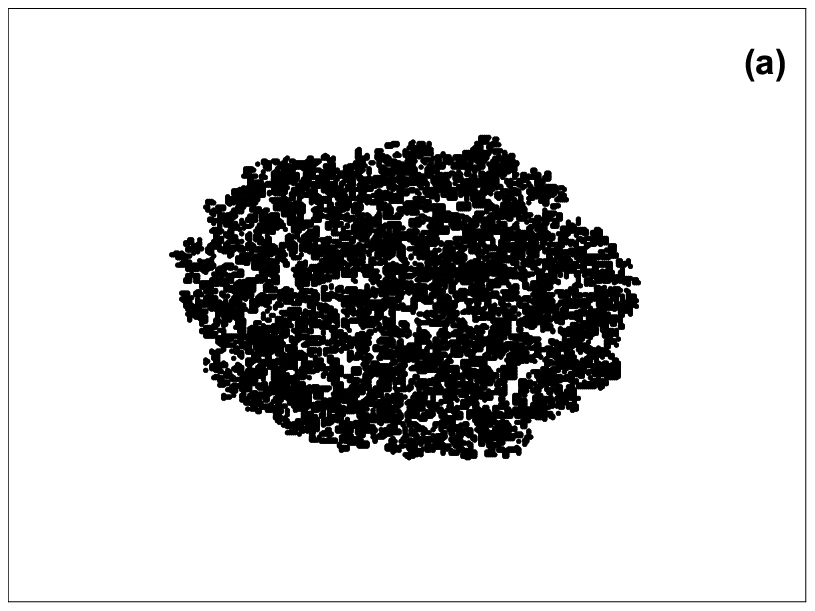}
\includegraphics[width=70mm,height=60mm]{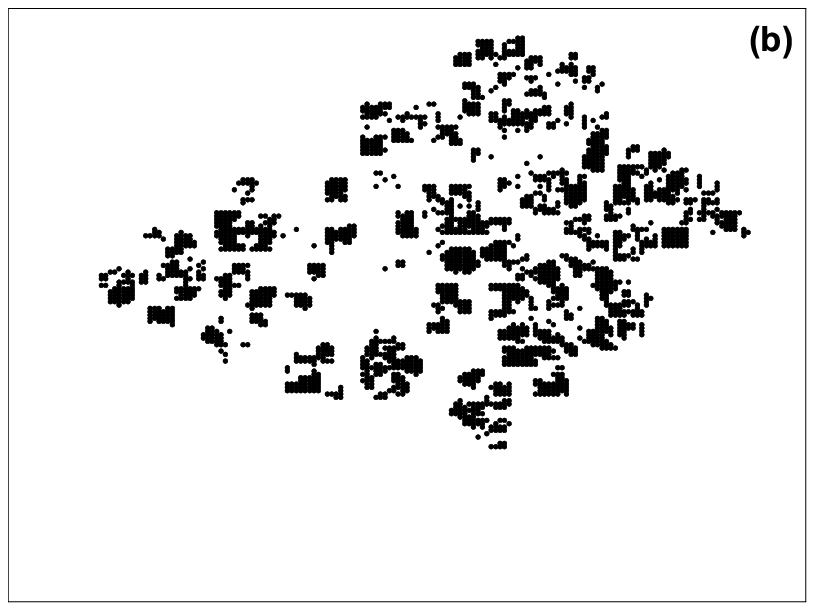}
\caption{\label{fig:12}(a) Snapshot of damaged lattice sites at $t=100$  
for $p=1.0$.(b) Snapshot of damaged lattice sites at $t=100$ 
for 
 $p=0.88$. The initial damage is of one site at 
 the center of $400\times400$ lattice.}
\end{figure}

\section{Conclusions}
  
We have studied the PD on a $2-D$ lattice with 
an update rule which
interpolates between asynchronous and synchronous update as
a function of the parameter $p$. 
Here each agent updates the strategy with probability $p$ at each timestep. 
We observe 
that, this system crosses from mixed (active) phase to all defector 
(absorbing) phase when we vary the parameter $p$.
We studied the time evolution of this system and  
found that this system exhibits long 
lived metastable state near the critical point. 
The order parameter of 
this system  shows a
clear jump at the critical point.  We carry out detailed 
quantitative analysis to show that the
the above transition s a first order transition.
We confirm this result by studying the average lifetime of metastable state,
Binder cumulant and hysteresis
effect.
   
The damage spreading technique is useful tool to study the sensitivity of the 
system dynamics on the initial condition. A damage spreading analysis of 
semi-synchronous update leads to the conclusion that, the active phase is 
chaotic and the transition in this system is as same  the frozen chaotic 
transition. The damage spreading inside the active phase (far from critical 
point) is compact. However, near the critical point the damage spreading is 
fractal. 
\section{Acknowledgments}
 MAS thanks Govt. of Yemen
for scholarship. PMG thanks DST for financial assistance.

%\end{center}
\end{document}